\documentclass[DIV12]{scrartcl}

\usepackage{hyperref}
\usepackage[utf8]{inputenc}
\usepackage{tabularx}
\usepackage{setspace}
\usepackage{graphicx}
\usepackage{booktabs}
\usepackage{amsmath}
\usepackage{subcaption}
\hypersetup{
	colorlinks=true,
	linkcolor=black,
	citecolor=black,
	urlcolor=black
}

\date{}
\usepackage{authblk}
\usepackage{blindtext}

\usepackage{natbib}
\title{Toward Inverse Control of \\ Physics-Based Sound Synthesis}

\usepackage{fancyhdr}
\begin{document}

\author[1]{A. Pfalz}
\author[1]{E. Berdahl}

\affil[1]{\small Louisiana State University}

\maketitle

\thispagestyle{fancy} 

\begin{abstract}
Long Short-Term Memory networks (LSTMs) can be trained to realize inverse control of physics-based sound synthesizers.  Physics-based sound synthesizers simulate the laws of physics to produce output sound according to input gesture signals.  When a user's gestures are measured in real time, she or he can use them to control physics-based sound synthesizers, thereby creating simulated virtual instruments.

An intriguing question is how to program a computer to learn to play such physics-based models. This work demonstrates that LSTMs can be trained to accomplish this inverse control task with four physics-based sound synthesizers.

\bigskip

\noindent {\textbf{Keywords:}} LSTM, physics-based models, sound synthesis

\end{abstract}

\section{Introduction}

\subsection{Sound Matching}
Controlling the parameters of sound synthesizers in order to realize target sounds has been a challenge for decades.  For instance, with the Frequency Modulation (FM) synthesis technique, the connection between the carrier frequencies, modulation frequencies, and modulation indices (particularly as these are automated) and the sound produced is complicated.  Accordingly, researchers have applied various techniques for adapting FM parameters to achieve target sounds such as \cite{horner1993machine}, \cite{garcia2001growing}, \cite{lai2006automated}, and \cite{tan1996automated}. 
Similar work as also been conducted for tuning the parameters of physics-based models; however, most of these works have required very specific optimizations that apply only to certain models (see \cite{sondhi1983inverse} and \cite{riionheimo2002parameter}).

\subsection{Inverse Control}
However, since humans are able to learn to play musical instruments, it seems very likely that machine learning methods could help address the inverse control problem.  For example, deep as well as shallow learning for audio and music generation has been investigated from a number of different perspectives. One approach is generating sequences of notes using MIDI or symbolic notation (as in \cite{magenta}). WaveNet is another very promising project among others (\cite{van2016wavenet}).  

The advantage of this technique is that it allows the generation of sounds to be automated. If a complex sound is desired, the user needs to spend time practicing to execute a potentially difficult gesture to be able to produce the output audio.

\begin{figure}
\centering
\begin{subfigure}{.38\textwidth}
  \centering
  \includegraphics[width=1\textwidth]{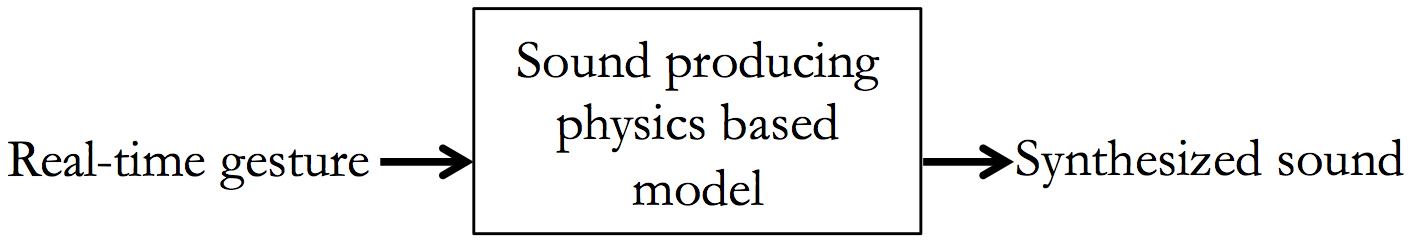}
\caption{Typical paradigm for physics-based sound synthesis (forward control).}
\label{fig:normal_usage}
\end{subfigure}%
\quad
\begin{subfigure}{0.53\textwidth}
  \centering
  \includegraphics[width=1\textwidth]{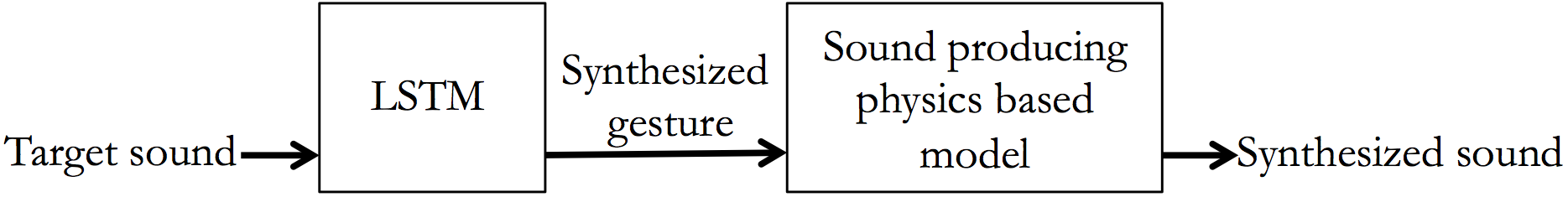}
\caption{Proposed paradigm (inverse control).}
\label{fig:new_usage}
\end{subfigure}
\caption{Paradigms for physics-based sound synthesis.}
\label{fig:Pluck}
\end{figure}

\section{Approach}

\subsection{Overview}
Synth-A-Modeler is an open-source software project for building physics-based sound synthesizers (see \cite{berdahl2012introduction}). Figure \ref{fig:normal_usage} shows the typical paradigm for  physics-based sound synthesis. In contrast, the present work proposes a new paradigm in which an LSTM learns to realize inverse control of a physics-based sound synthesizer.  In other words, the LSTM uses a \emph{target sound} to match the \emph{gesture} that was used to create the \emph{target sound} using a physics-based sound synthesizer.  Consequently, the predicted \emph{synthesized gesture} can be applied to a physics-based sound synthesizer to create a \emph{synthesized sound}.  Although not studied directly in this introductory work, this paradigm has potential applications denoising audio recordings, toolkits for making foley, sound transformations and sound mappings.

\subsection{Physics-Based Sound Synthesizers Used in the Project}
\begin{figure}
\centering
\begin{subfigure}{.5\textwidth}
  \centering
  \includegraphics[width=1\linewidth]{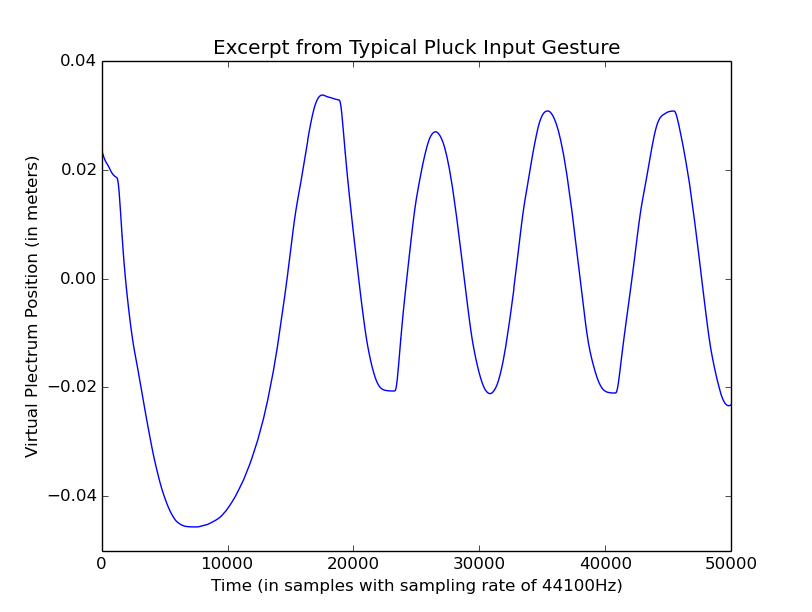}
  \caption{Gesture signal.}
  \label{fig:pluck_input}
\end{subfigure}%
\begin{subfigure}{.5\textwidth}
  \centering
  \includegraphics[width=1\linewidth]{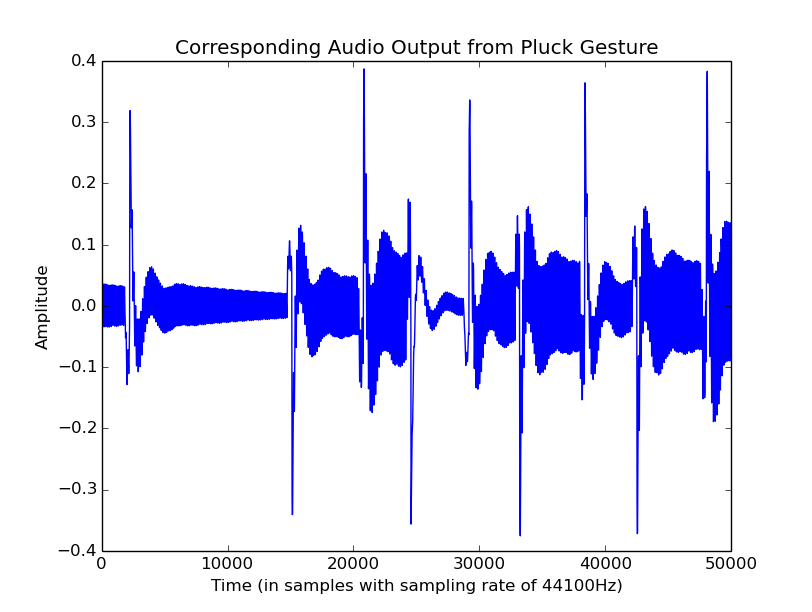}
  \caption{Synthesized sound output.}
  \label{fig:pluck_output}
\end{subfigure}
\caption{Gesture signal input and corresponding synthesized sound output for a physics-based model with a plucking excitation.  (The horizontal axis is time in samples.)}
\label{fig:Pluck}
\end{figure}

Each physics-based sound synthesizer used in this work receives a gesture signal that could for example represent the position of a musician's hand in real-time, enabling her or him to play virtual sound synthesizers. These gesture signals are in the range of $-0.05\ \mathrm{m}$ to $0.05\ \mathrm{m}$ (see Figure \ref{fig:pluck_input} for an example gesture) with an audio sampling rate of 44.1kHz. Depending on the particular sound synthesizer type, the gesture excites the synthesizer in a different way. 

For example, the sound synthesizer \texttt{PluckAResonator.mdl} incorporates a virtual plucking mechanism that activates a single, virtually oscillating resonator.  Accordingly, the sound is triggered when the plectrum is pushed beyond the $0 \ \mathrm{m}$ point (see Figure \ref{fig:Pluck}).\footnote{To model the dynamics of a plectrum more precisely, the sound is actually triggered when the plectrum is pushed a small distance $d$ meters beyond the \emph{difference} between the virtual gesture input signal and the current position of the resonator, where $d$ changes sign with each pluck.  The details of how the sound synthesizers work are beyond the scope of this paper. Their dynamical behaviors are complex, nonlinear and nuanced, as is appropriate for modern physics-based sound synthesis.}
 
For example, as the gesture signal moves up from $-0.05\ \mathrm{m}$, there will at first be no sound synthesized until the virtual plectrum moves close enough to interact with the virtual resonator (see Figure \ref{fig:Pluck}). Then, as the gesture continues moving up, the plectrum ``plucks'' the virtual resonator causing it to vibrate and produce sound (see Figure \ref{fig:Pluck} near $t=16000$ samples). Notice that the spikes in amplitude in the synthesized sound signal correspond approximately to zero crossings in the gesture signal. 

Different nonlinear excitation mechanisms were employed in the models described below, to verify that the LSTMs could, just like human musicians, learn to produce a wide range of gestures for this application.   For example, the synthesizer \texttt{TouchSeveralModalResn.mdl} was controlled using a  kind of ``nonlinear contact link,''   \texttt{ScratchMassLinkChain.mdl} incorporated a bowed-string kind of interaction, and \texttt{PluckHarp10.mdl} featured ten individually pluckable virtual strings, each with its own distinct plucking point across the input gesture range.

\subsection{An LSTM Network for Inverse Control}
Long Short Term Memory (LSTM) networks are known for their sequence prediction abilities (see \cite{Hochreiter:1997:LSM:1246443.1246450}).  In this work, it was decided to try using LSTMs for inverse control in order to see if an LSTM could learn to play music using physics-based sound synthesizers.

\begin{figure}
\centering
\includegraphics[width=.7\linewidth]{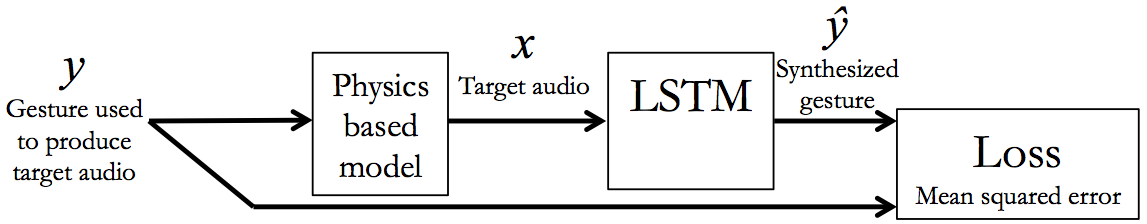}
\caption{Diagram of how loss is computed.}
\label{fig:loss}
 \end{figure}

LSTMs were implemented using the high level TensorFlow API \texttt{contrib.rnn}. The models had two or three layers each with 1024 units implemented with \texttt{contrib.rnn.MultiRNNCell} wrapper with default settings except where indicated otherwise (\cite{tensorflow2015-whitepaper}). The loss function used the mean-squared error to measure the similarity between a synthesized gesture $\hat{y}$ and a gesture $y$ that was used to produce a target sound $x$. The Adam Optimizer was used.  Figure \ref{fig:loss} shows the process of generating the data and training an LSTM.

In order to capture the most data with a single input, the audio data in $x$ was downsampled from 44100Hz to 2756Hz (a speech-quality sampling rate) before showing it to the LSTM. To obtain gesture signals, six-minute recordings were made with each physics-based sound synthesizer of a human musician performing musical gestures using a single degree-of-freedom haptic device (\cite{berdahl2012firefader}).  This recording was then broken into segments that were 1024 samples long. Batches of 98 inputs were used per training iteration. The models were trained over 64 epochs. The validation and testing datasets were each 10\% of the  original six-minute corpus.

\section{Results}

Audio examples of results from these experiments are available at \url{https://cct.lsu.edu/~apfalz/inverse_control.html}. In each of the files, the left channel is the synthesized gesture $\hat{y}$. The right channel is the target audio $x$. All the examples come from the test set.

Audio example 1 shows the simplest of the physics based models. The resultant audio matches the target audio quite closely. There are slight differences in the amplitude and phase on the audio. The authors found the differences are barely noticeable and only when listening on headphones. Audio example 2 shows comparable results with the physics based model that uses a virtual touch for its mode of interaction.

Audio example 3 shows the most impressive results from these experiments. In this example, the LSTM had to learn to consider the frequencies of the target audio it was trying to match. In some informal trials not shown here, the LSTM was shown to be able to predict comparably accurate gestures for simpler physics based models like the one shown in audio example 1 when the LSTM was shown only the RMS level of the target audio, rather than the raw audio itself. 

Audio example 4 shows the worst performance. This was expected because the target audio is decidedly more complex than the other inputs. The LSTM was still able to predict a gesture that closely matched the target audio. However, there the gesture contained some extraneous excitations.

The normalized absolute error for the test set was calculated using equation \ref{eqn:error}.

\begin{equation}
Normalized Absolute Error = \frac{\frac{1}{N} \sum^N_{i=1}|Y_i - \hat{Y}_i|}{\frac{1}{N}\sum^N_{i=1}|\hat{Y}_i|}, 
\label{eqn:error}
\end{equation}
where $Y$ is the targets and $\hat{Y}$ is the prediction. The model was able to synthesize the target gesture very accurately for \texttt{TouchSeveralModalResn.mdl} and \texttt{PluckHarp10.mdl} with 4.12\% and 2.20\% error respectively. \texttt{TouchSeveralModalResn.mdl} and
\texttt{ScratchMassLinkChain.mdl} performed worse with 22.37\% and 16.84\% error respectively. The audio that resulted from these gestures still matched the target audio closely though.


\section{Conclusions and Future Work}

The LSTM was able to synthesize gestures for inverse control of physics-based sound synthesis. The synthesized gestures could be used with physics-based models for re-synthesizing audio that, at least in the opinion of the authors, subjectively resembled the target audio quite closely, the authors hope that readers will visit the project web page and listen to the sound examples to judge for themselves.  Moreover, the gestures synthesized by the LSTM matched the target gestures. Applications of this technology could include denoising of audio recordings, toolkits for making foley, as well as new kinds of sound transformations and mappings, which can be achieved by applying the synthesized gesture signals to a diverse range of physics-based sound synthesizers.
 
More generality might come from training in such a way that the LSTM produces a synthesized gesture that would produce match the target audio but without matching the particular gesture that created the input to LSTM. Measuring the loss against the audio directly rather than against the gestures would remedy this. The physics based models are capable of a wider range of sounds than are demonstrated in the audio inputs presented here. It should be investigated to what extent an LSTM could model both the input gesture and time-varying parameters like fundamental frequency or amplitude. Also larger datasets can be generated randomly instead of only using input from a human.

\bibliography{paper}

\end{document}